\documentclass[11pt]{article}

\usepackage[parsep]{collref}

\usepackage{latexsym,amsmath,amssymb,theorem,epsfig}
\usepackage{amsmath}

\DeclareFontFamily{OT1}{rsfs10}{}
\DeclareFontShape{OT1}{rsfs10}{m}{n}{ <-> rsfs10 }{}
\DeclareMathAlphabet{\mathscript}{OT1}{rsfs10}{m}{n}

\numberwithin{equation}{section}

\usepackage[T1]{fontenc}
\usepackage[usenames,dvipsnames]{xcolor}
\usepackage[setpagesize=false,pagebackref=false, 
linktocpage, bookmarksopen=true, colorlinks=true, 
linkcolor=blue,citecolor=blue,urlcolor=blue]{hyperref}
\usepackage{tikz} 
\usetikzlibrary{arrows,decorations.pathreplacing,decorations.markings,snakes}
\usetikzlibrary{cd}

\newcommand{\RR}{{\mathbf{\rR}}}

\newcommand{\com}[2]{[#1,#2]}

\def\a{\alpha}

\def\g{\gamma}

\def\l{\lambda}
\def\m{\mu}
\def\n{\nu}

\def\s{\sigma}
\def\t{\tau}

\def\G{\Gamma}

\def\P{\Pi}

\def \S {{\cal S}}

\def\gsim{ \lower .75ex \hbox{$\sim$} \llap{\raise .27ex \hbox{$>$}} }
\def\lsim{ \lower .75ex \hbox{$\sim$} \llap{\raise .27ex \hbox{$<$}} }
\def\be{\begin{equation}}
\def\ee{\end{equation}}
\def\bea{\begin{eqnarray}}
\def\eea{\end{eqnarray}}

\def \td {\tilde}

\def \ha {{1 \ov 2}}

\def \del{\partial}
\def \a {\alpha}

\def\ov{\over}
\def \ci {\cite}

\def \foot {\footnote}

\def\la{\label}
\def\foot{\footnote}
\newcommand{\rf}[1]{(\ref{#1})}
\def \OO {{\cal  O}}\def \no {\nonumber}

\def \LL {{\rm L}}

\def \g {{\gamma} } \def \hD {\hat \D}
\def \ov {\over}

\def \n {\nu}

\def \Tr {{\rm Tr}}
\def \iffa {\iffalse}

\usepackage{tikz}
\usetikzlibrary{decorations.markings}

\makeatletter
\renewcommand*{\@fnsymbol}[1]{\textit{\@alph{#1}}}
\makeatother

\begin{document}
\begin{titlepage}
\vspace{-15cm}
\vspace{-4cm}
\title{\vspace{-3cm}
   \vspace{1cm} 
   { 
   \bf On world-sheet S-matrix \\[0.02em]
   of NSR string in static gauge  }
  \\[1em] }

\author{\large    Arkady A. Tseytlin\footnote{Also at  the Institute  for Theoretical and Mathematical Physics    (ITMP)  of  MSU
      and  Lebedev
    Institute.}  \ \ \ \ \ 
    and \ \ \ \ Zihan Wang\thanks{zihan.wang18@imperial.ac.uk} 
       \\[0.9em]
\it      Abdus Salam Centre for Theoretical Physics \\[0.03em]
  \it   Imperial College London,  SW7 2AZ, UK 
}

  \maketitle

\begin{abstract}
As was shown  in arXiv:1203.1054,  expanding the Nambu action near the "long string" vacuum  in the static gauge   one 
 finds  that  the  one-loop $2 \to 2$ scattering  amplitude of  the $D-2=24$  transverse 2d  fluctuations  $X^i$  is given  by a pure phase expression  consistent with underlying integrability. 
 Similar computation in the Green-Schwarz   superstring  was carried out in arXiv:2404.09658. 
 Here we consider the case of the NSR or spinning   string   starting with its  manifestly 2d  covariant  action 
 given by $D$  scalar multiplets   coupled to 2d supergravity.   Integrating out  the  zweibein and gravitino  fields 
 is not straightforward    so that a  Nambu-like  version of the  spinning string action
  involving only  the scalar coordinates  and their 2d fermionic partners   was not  given  in the past. 
 We show  how  the elimination of the auxiliary   fields   can be done  in the static gauge  after a particular choice of the 
 superconformal gauges on the fermions   while   keeping  the  transverse fields $X^i$ and $\psi^i$ off-shell. 
 The  resulting one-loop S-matrix   for $X^i$   turns out, as expected, to be the same as in the GS superstring case. 
 We also discuss  similar  actions  for  the heterotic string and  a  $T\bar T$ deformation of a   free  2d scalar multiplet. 
 \end{abstract}

 \iffa 
 As was shown  in arXiv:1203.1054,  expanding the Nambu action near the "long string" vacuum  in the static gauge   one 
 finds  that  the  one-loop 2 to 2 scattering  amplitude of  the $D-2=24$  transverse 2d  fluctuations  $X^i$  is given  by a pure phase expression  consistent with underlying integrability.  Similar computation in the Green-Schwarz   superstring  was carried out in arXiv:2404.09658. Here we consider the case of the NSR (or spinning)   string   starting with its  manifestly 2d  covariant  action given by $D$  scalar multiplets   coupled to 2d supergravity.  Eliminating the zweibein and gravitino fields is non-trivial, and a Nambu-like formulation of the spinning string Ñ involving only scalar coordinates and their 2d fermionic partners Ñ has not previously been available. We show how the auxiliary fields can be eliminated in static gauge after a specific choice of superconformal gauge for the fermions, while keeping the transverse fields $X^i$ and $\psi^i$ off-shell. The  resulting one-loop S-matrix for $X^i$   is found, as expected, to be the same as in the GS superstring case. We also discuss  similar  actions  for  the heterotic string and  a  $T\bar T$ deformation of free  2d scalar multiplet. 
 \fi
 
%
\end{titlepage}

\def \iffa  {\iffalse}
\def \RR {{\mathbb R}}
\def \R {{\rm R}}
\def \s {\sigma} \def \t {\tau} 
\def \G {\Gamma} \def \four {{1\ov 4}}
\def \CP  {{\rm CP}}\def \gs {g_s}
\def \hD   {\hat{D}}
\def \JJ {{\rm J}}
\def \te {\textstyle}
\def \zz  {^{(0)}} 
\def \ve {\varepsilon}
\def \zo {^{(1)}}
 \def \rR  {{\rm R}} 
 \def \eps {\epsilon}
 \def \rS  {{\rm S}} \def \rT {{\rm T}}
\def \P   {{\rm P}}
 \def \bbeta {{\rm r}} 
\def \no {\nonumber}
\def \TTT  {{\cal T}} 
\def \S {{\rm S}}  \def \rT  {{\cal T}} 
\def \dDelta {{\hat \Delta}}
\def \ss {\tau}
\def \tn {\lambda} 
\def \hD {\hat D}
\def \half {\tfrac{1}{2}}

\def \ba {\begin{align}}
\def \ea {\end{align}}
\def \RR  {{\mathbb R}}
\def \rr {{\rm R}}  \def \bQ {{\bar Q}}
\def \nf {{\rm n}_{_{\rm F}}}
\def \hT  {\hat T}

 \def \bD {{\rm  b}}  
   \def \rD {{\rm D}} 
\def \LL {{\rm L}}
\def \vt {\vartheta}
\def \ln {{\rm log\,}}
\def \tad {{\rm tad}}
\def \ed {
\small
\bibliography{biblio2.bib}
\bibliographystyle{JHEP-v2.9}
\end{document}
}
\def \edo {\end{document}}

\def \com {} 
\def \un {\underline}
\def \vv {{\rm v}}
\def \umu {\underline \mu}
\def \unu {\underline \nu}
\def \vp  {\varphi }  \def \rr {{\rm r}} 
\def \I {{\mathcal{I}}}  \def \j {{\mathcal{J}}}  

\tableofcontents

\section{Introduction}


Starting with the Nambu action in flat $D$-dimensional 
space-time   one may expand it near  the "long string" vacuum $X^a=\xi^a$ ($a=0,1$). 
In the static gauge  one gets  an  interacting action for the $D-2$ physical 2d massless  scalars $X^i$   
\be \la{1}
S= -T \int d^2 \xi \, \sqrt{ -\det ( \eta_{ab} + \del_a X^i \del_b X^i) } 
= -T \int d^2 \xi \,  \big[1 + {1\ov 2} \del_a X^i \del^a X^i + \OO ((\del X)^4) \big] \ . \ee
The resulting 4-point scattering amplitude of $X^i$ is straightforward to compute 
at  the tree  and one-loop  \ci{Dubovsky:2012sh,Dubovsky:2012wk}  (and 2-loop \ci{Conkey:2016qju}) 
orders and it is  found to be consistent   with integrability of the underlying theory  for $D=26$.

Similar  one-loop  computation of the bosonic amplitude 
  can be done in the $D=10$ Green-Schwarz   superstring case \ci{Seibold:2024oyr}
  (cf.  also \ci{Cooper:2014noa,Mohsen:2016lch}). This   is done  by starting  with the  GS   generalization of the 
  Nambu action (found  by  solving for the independent 2d metric $g_{ab}$)  and  then  fixing the static gauge on
  bosonic coordinates  $X^A=(\xi^a, X^i(\xi))$  
  and an appropriate $\kappa$-symmetry gauge 
  on  their   fermionic  counterparts $\theta$.
    The  resulting 
  action expanded  in  terms of 8 commuting  and 16 real anticommuting 
     physical components of the fields can be used  to compute the 4-scalar   scattering 
   amplitude  containing   the fermionic loop contribution.

   One may wonder how an analogous    computation of S-matrix  in the   long string  vacuum 
    can be done  in the case of  the NSR  or spinning  string
    that has manifest world-sheet supersymmetry. 
    The  corresponding  bosonic amplitude  is expected  to  be  the same  as  the GS  string 
    one in $D=10$ (two theories are equivalent  in the light-cone gauge  \ci{Witten:1983ymv,Green:2012oqa}).
     However,  it  turns out to be   non-trivial to demonstrate this  by starting  directly 
    with  the  covariant  spinning   string action  found in 
    \ci{Brink:1976sc,Deser:1976rb}.
     The corresponding Lagrangian  describes $D$ copies   of  2d real scalar multiplet 
    $(X^A, \psi^A)$   coupled to 2d $N=1$ conformal  supergravity fields $(e^a_\mu, \chi_\mu)$
  \begin{align}
\label{1.2}
\mathcal{L}= &  e\Big[-\frac{1}{2}\partial_{\mu}X^{A}\partial^{\mu}X_{A}-\frac{i}{2}\bar{\psi}^A\gamma^{\mu}\partial_{\mu}\psi_{A}+\frac{i}{2}\bar{\chi}_{\mu}\gamma^{\nu}\gamma^{\mu}\psi^{A}\partial_{\nu}X_{A}-\frac{1}{16}\bar{\chi}_\mu \gamma^\nu \gamma^\mu \chi_\nu \bar{\psi}^A\psi_A\Big],
\end{align}
where $\gamma^\mu = e_a^\mu \gamma^a\,, \  \{\gamma^a, \gamma^b\} = 2\eta^{ab}$.
A reparametrization-covariant Nambu-like  counterpart  of this  action  involving only $X^A$ and $\psi^A$ 
was not  discussed  in the past:  given a  non-linear dependence  of \rf{1.2} on  the gravitino  $\chi_\mu$
 is  not   immediately clear   how to  integrate  it out     and then solve for $e^a_\mu$. 
 
 Fixing the static gauge $X^a=\xi^a$,  to compute the scattering amplitude for $X^i$ 
  one needs an   action containing  only the physical components $(X^i,\psi^i)$
  that should  not be constrained  by any additional conditions. 
  A   seemingly natural  approach is   to use the local superconformal symmetry to gauge-fix $\chi_\mu=0$ 
 and then try to solve the constraint   resulting from  variation over $\chi_\mu$   to eliminate  the longitudinal components $\psi^a= (\psi^0,\psi^1)$  but this appears to be hard to implement explicitly. 
 
 Our aim below  will be to show   how this  problem  
 can be overcome  using a  different gauge fixing procedure.\foot{The computation of one-loop 
 scattering amplitude for $X^i$ in  the   NSR case was  previously discussed   in \ci{Mohsen:2016lch}
 using  an  action for $(X^i,\psi^i)$   found   by requiring a   non-linear realization 
 of the underlying  $D$-dimensional Poincare symmetry. This approach did not start with a fully covariant  spinning 
 string    action  and 
 it appears to be hard to verify its   validity (e.g. the uniqueness of the resulting action).}
 This   will be the topic of section 2.
 In section 3 we will discuss  similar  static-gauge actions  for the heterotic string and GS superstring. 
  In section 4  we will  compute the   spinning string  one-loop scattering amplitude for 4 
   massless scalars $X^i$  and show that the result is the same as in the superstring case  in \ci{Seibold:2024oyr}.

  One motivation   to  study the action of the  spinning string in the static gauge 
  is  to compare  it  to a   $T\bar{T}$ deformation  of  the free 2d scalar multiplet 
  $(X^i, \psi^i)$    (cf.  \ci{Baggio:2018rpv,Chang:2018dge,Jiang:2019hux,Frolov:2019nrr, Coleman:2019dvf,Lee:2023uxj,Tsolakidis:2024wut}). We 
   will comment  on this  in section 2.4 below.

 Both  the  Nambu  and the GS   string  world-sheet S-matrix computations can be  generalized 
  to the  bosonic membrane   and  space-time supersymmetric  M2-brane   \ci{Bergshoeff:1987cm}  cases
   \ci{Seibold:2023zkz,Seibold:2024oyr}.
  One may wonder  if   a similar computation can be done 
  for a 3d world-volume supersymmetric  analog of the    bosonic membrane action.
     Like in the spinning string case, 
    to  have  3d reparametrization invariance  and  supersymmetry one is to couple a  collection of $D$ 3d  real 
    scalar  multiplets
  $(X^A, \psi^A)$ to a  3d supergravity multiplet. 
   Attempts to construct such 3d supersymmetric membrane  action    were discussed in  \cite{Howe:1977hp,Howe:1977us,Howe:1977ut}.  In contrast to the 2d  spinning string case where 
   the Einstein term in the supergravity action is a total derivative   
    this is not  so  in the 3d  case  and a  consistent 3d   supersymmetric analog of  the  membrane action 
    appears to require  the inclusion of the 3d supergravity part 
   (cf. \cite{Bergshoeff:1988ui, Luckock:1989jr}). In this  case the metric   and gravitino  become dynamical   and their  contributions   should be properly accounted for when  computing the S-matrix for the   bosonic $X^i$ 
   fields. We hope to discuss this  computation in the future.
   

\iffa
     Another question is the higher derivative terms of the supermembrane action.
In the bosonic long string, there could be higher-derivative terms dependent on extrinsic curvature \cite{Polyakov:1986cs,Kleinert:1986bk}. Constructing a general action invariant under non-linearly realized spacetime PoincarÃ© symmetry $ISO(1,D-1)$ gives the Nambu-Goto action plus a rigidity term \cite{Dubovsky_2012}. The rigidity term does not contribute to the scattering amplitude at tree-level and one-loop level, the scattering is dependent on the Nambu-Goto part of the action. At one-loop level, the scattering amplitude is found to have logarithmic divergence and requires a counterterm. However, there is no tree-level vertices contributing to the $2\rightarrow2$ scattering with the same $s^3$ dependence. This is resolved by adding an Einstein-Hilbert term to the action, which does not change the tree-level scattering amplitude in two dimensions but has to be included in the dimensional regularization to remove divergences. Similarly, we could consider higher derivative terms for the supermembrane  \cite{Howe_2003,Howe:2001wc}.
\fi

\section{Spinning string  action  expanded near long string vacuum}

To compute the  world-sheet S-matrix  in  the long string vacuum 
the first step is  to expand the action \rf{1.2}   in the static gauge. This requires  gauge fixing all  local symmetries
and integrating out the auxiliary fields    to end up with an action for the physical $D-2$ components 
of the bosonic and fermionic fields $(X^i, \psi^i)$.

\subsection{Gauge fixing}

Let us first recall the notation we used   in the spinning string action \rf{1.2}  (see also Appendix A).
 The world-sheet  coordinates are $\xi^a$  ($a=0,1$), 
the  induced metric is $g_{\mu \nu} = e_\mu^a e_\nu^b \eta_{ab}, \  \eta_{ab}={\rm diag}(-1,1)$, \  ($\mu,\nu=0,1$).
The index $A=0,...,D-1$ labels the flat target space coordinates. 
$\psi^A$ are  2d  Majorana fermions in real representation  with 
$\bar{\psi}^A=(\psi^A)^T \gamma^0$,  and  constant   flat-space Dirac matrices $\g^a$ are related to 
$\g^\m$ in \rf{1.2}  by $e^a_\m$.  We will denote the   spinor components of $\psi$ as 
{\tiny $\begin{pmatrix}
\psi_R \\
\psi_L
\end{pmatrix}$},  with $\bar{\psi}= (\psi_L, - \psi_R)$  and similarly for $\chi_\m$. 

In addition to the reparametrization invariance
 and  local Lorentz  rotations given by 
 \begin{equation}
\label{21}
    \delta \psi^A=\frac{1}{2}\alpha \gamma^*\psi^A, \qquad
     \delta e^a_\mu=\alpha \epsilon^a{}_b e^b_\mu, \qquad 
      \delta \chi_\mu=\frac{1}{2}\alpha \gamma^* \chi_\mu\,, \qquad 
       \gamma^* \equiv  \ha \epsilon_{ab} \g^a \g^b =\g^0\g^1 \ , 
\end{equation}
  the  action \rf{1.2} is invariant under the  local supersymmetry transformations (see  \ci{Brink:1976sc,Deser:1976rb} for details)
\begin{equation}
\label{23}
    \delta X^A = i \bar{\varepsilon} \psi^A\,, \qquad \delta \psi^A = \gamma^\mu (\partial_\mu X^A - \frac{i}{2} \bar{\chi}_\mu \psi^A) \varepsilon\,, \qquad \delta e^a_\mu={i}\bar{\varepsilon}\gamma^a\chi_\mu\,,\qquad \delta \chi_\mu=2 \nabla_\mu \varepsilon\,, 
\end{equation}
 and also the   Weyl  and super-Weyl  transformations
\begin{equation}
\label{24}
    \delta X^A=0\,, \quad \delta e^a_\mu=\lambda e^a_\mu\,, \quad \delta \psi^A =-\frac{1}{2}\lambda \psi^A\,, \quad \delta \chi_\mu =\frac{1}{2}\lambda \chi_\mu,\qquad \qquad 
    \delta \chi_\mu =\gamma_\mu \eta\,.
\end{equation}
We shall  fix the reparametrizations  by choosing the static gauge, i.e. 
setting  the fluctuations  of $X^a$ to zero, 
\begin{equation}
    X^A=(\xi^a,X^i),\qquad a=0,1,\ \ \ \qquad i=2,3,...,D-1.
\end{equation}
Then the variation under the  local  supersymmetry  \rf{23} of the
 first two components of $\psi^A$  takes the form 
  $\delta \psi^a = \gamma^\mu  \delta_\mu ^a \varepsilon  - \frac{i}{2} \bar{\chi}_\mu \psi^a \varepsilon$.
  The presence of the  first  inhomogeneous term   implies that half of the $\psi^a$ components 
   are  Stueckelberg fields   so that they can be  eliminated by a gauge fixing.\foot{This parallels the
  logic  that  underlines the  bosonic static gauge  fixing: 
  under  the reparametrizations $\delta X^a = \zeta^\m \del_\m X^a$ 
  so that for $X^a = \xi^a + \td X^a$  one gets $\delta \td X^a =\zeta^\m \delta^a_\m  +  \zeta^\m \del_\m \td X^a$. This  
  implies  that $\td X^a$  is a Stueckelberg field   that can be gauge-fixed to zero.}
  Explicitly, we  shall 
   fix the local supersymmetry  by the following gauge\foot{From \rf{23} one has  $ \delta(\gamma^a \psi_a)= M \varepsilon$
   where  $M\equiv\gamma^a \gamma^\rho\big(\eta_{\rho a}-\frac{i}{2} \bar{\chi}_\rho \psi_a\big)$ is a diagonal 2 by 2 
   matrix in spinor components  which is  thus  invertible,  allowing one to  find an $\varepsilon$ such that \rf{25} is satisfied.}
  \begin{equation}
\label{25}
      \gamma_a \psi^a=0, \qquad \ \ \ \ i.e.\ \ \ \ \qquad \psi_R^{-}= 0, \ \ \  \psi_L^{+}=0, \ \ \ \ \ \ \qquad \psi^\pm \equiv \psi^0 \pm \psi^1 \ . 
\end{equation}
An important  consequence of \rf{25}   is that  
  the kinetic term for $\psi^a$  in the action \rf{1.2}  then vanishes (see (\ref{g0d}) and (\ref{g1d}))
\begin{equation}
\begin{aligned}\la{2.6}
\bar{\psi}^{a}\gamma^{\mu}\partial_{\mu}\psi_{a}&=e^\mu_b\bar{\psi}^a\gamma^b\partial_\mu \psi_a=0.
\end{aligned}
\end{equation}
  The  super-Weyl  transformation of the gravitino  in \rf{24}  implies  that $\gamma^\mu \chi_\mu$ 
  is another Stueckelberg field  that 
  can be fixed   by the gauge 
\begin{equation}
\label{28}
    \gamma^\mu \chi_\mu=0 \ .  
\end{equation}
As a result,  the   Lagrangian  in \rf{1.2}  takes  the form 
\begin{equation}
\label{L2}
\begin{aligned}
     \mathcal{L}=e\Big[-\frac{1}{2} & (\eta_{\mu \nu}+ \partial_\mu X^i \partial_\nu X^i) g^{\m\n}  -\frac{i}{2} \bar{\psi}^i \gamma^\mu \partial_\mu \psi^i+{i} \bar{\chi}^\mu  \psi_{\underline{\mu}} +{i} \bar{\chi}^\mu \psi^i \partial_\mu X^i-\frac{1}{8}\bar{\chi}^\mu\chi_\mu \bar{\psi}^A\psi_A\Big],
\end{aligned}
\end{equation}
where we introduced the notation 
\be  \eta_{\mu \nu}\equiv  \delta^a_\mu \delta^b_\nu \eta_{ab}  \ , \qquad \qquad     \psi_{\underline{\mu}}\equiv \delta^a_\mu \psi_a\ . \la{30} \ee 
  After imposing  the gauge  conditions \rf{25} and \rf{28}   we are left with  2+2    independent  spinor components  of 
   the  fermionic  fields $\psi^a$ and 
  $\chi_\mu$. These enter \rf{L2} without derivatives and thus  play the role of auxiliary fields 
   that  can  be integrated out  algebraically.

 Similar decomposition into    Stueckelberg parts (that  can be gauge-fixed) 
    and auxiliary  parts (that should be integrated out)   applies to  the zweibein.   As follows from \rf{21} and \rf{24} 
     the Weyl factor and antisymmetric 
    part of $e_{a\m}\equiv \eta_{ab} e^b_\mu $  can be  gauge-fixed, e.g., by setting $e_{a\mu}$ to be symmetric matrix with determinant  one, i.e.\foot{As was noted in \ci{Deser:1976rb}, 
    by fixing the gauges  $\gamma^\mu \chi_\mu=0$, $e^0_1=0$ and $e=1$ 
     the action (\ref{1.2}) becomes   equivalent to the one  constructed in \cite{Collins:1976fp,Collins:1976yy}.} 
    \begin{equation}
\label{ge}  e_{[a\mu]}=0,   \ \ \qquad  \ \ \  \quad e\equiv \det e^a_\mu  =1  \ , 
\end{equation}
    while the remaining 2 independent components of $e^a_\m$ 
    should be integrated out.\foot{ In general, 
     the   strategy   should be  first  (i)     gauge-fixing  Stueckelberg   fields 
    and then (ii)  integrating out (or solving for)  the remaining auxiliary   fields.
    As Stueckelberg   fields transform  via  algebraic shifts   by gauge parameters, there are no non-trivial ghost determinants
    associated with setting them to zero. }

\subsection{Solving for the  auxiliary fields} 

Let us first  consider  eliminating   the remaining components of $\psi^a$ and 
  $\chi_\mu$. 
  Since these auxiliary fields appear in the Lagrangian \rf{L2} without  derivatives, the
  result of  integrating them out is an  "ultralocal"  contribution $\sim \delta^{(2)}(0)$
  that can be  absorbed into the path integral measure or  regularized away (it vanishes
   in dimensional regularization that we will use below). 
   
    Indeed,   viewing  the 4-fermion  term  $\bar{\chi}^\mu\chi_\mu \bar{\psi}^A\psi_A$ 
  as an interaction vertex, one   may use perturbation theory  by  expanding  in powers of this vertex
  near the quadratic  fermionic  part  given by  $\chi \psi$ terms in \rf{L2}. The
  contribution of the  quadratic  fermionic terms   produces a local determinant 
    while the "Feynman diagrams" generated by the quartic  fermionic  vertex 
  will  also  be proportional to $\delta^{(2)}(0)$.  As a result,  we are left  just with the   $\chi$-independent part of \rf{L2}, i.e. 
  ($g^{\m\n}= e^\m_a e^\n_b  \eta^{ab}$)
\begin{equation}
\label{L3}
\begin{aligned}
     \mathcal{L}=-\frac{1}{2} e\Big[ & (\eta_{\mu \nu}+  \partial_\mu X^i \partial_\nu X^i ) g^{\mu \nu}+ 
     i  e^\mu_a \bar{\psi}^i \gamma^a \partial_\mu \psi^i \Big] \ . 
\end{aligned}
\end{equation}
The same    conclusion is reached also   by just solving the non-linear   classical equations  for  $\chi_\m$  and $\psi^a$  
that follow  from \rf{L2}. 
Taking variations of \rf{L2} over $\chi_\mu$   and $\psi_a$   subject to the gauge conditions \rf{28} and \rf{25} 
we get (we use the same  notation as  in  \rf{30}) 
\begin{align}
\label{212}
  i\gamma^\nu \gamma^\mu \psi_{\underline{\nu}}&+ i \gamma^\nu \gamma^\mu \psi^i \partial_\nu X^i -\frac{1}{2} \chi^\mu \bar{\psi}^A \psi_A=0,
\\
\label{213}
&   i\gamma_{\underline \rho} \gamma^{a}\chi^\rho-\frac{1}{2}(\bar{\chi}_\mu \chi^\mu )\psi^a=0 \ . 
\end{align}
Eq. (\ref{213})  implies that (cf. \rf{25})
\begin{equation}
\label{215}
    \begin{aligned}
        &\chi^+_{R}=-\frac{i}{4}(\bar{\chi}^\mu \chi_\mu)\psi^+_R\ , \quad\qquad \qquad  \chi^-_{L}=-\frac{i}{4}(\bar{\chi}^\mu \chi_\mu)\psi^-_L,
    \end{aligned}
\end{equation}
where $\chi^\pm =\chi^0\pm \chi^1$  with  $\chi^\mu=(\chi^0,\chi^1$)   and 
$L$ and $R$ label the upper  and lower spinor components.
The gauge condition \rf{28}  implies that (here $e^0_0$, etc.,  are the components of  $e^a_\m$)
\begin{align}
\label{555}
 &   \chi^-_{R}=- \frac{e^0_0+e^0_1-e^1_0-e^1_1}{e^0_0-e^0_1-e^1_0+e^1_1}\chi^+_{R}\ , \quad\qquad \qquad  \chi^+_{L}=-\frac{e^0_0-e^0_1+e^1_0-e^1_1}{e^0_0+e^0_1+e^1_0+e^1_1}\chi^-_{L}\ , 
\\ \la{216}
 &   \bar{\chi}^\mu \chi_\mu=\, F(e)\, \chi^-_{L}\chi^+_{R} \ , \qquad \qquad \qquad \ \ 
    F(e) \equiv  -\frac{4e^2}{[(e^0_0+e^1_1)^2-(e^0_1+e^1_0)^2]} \ . 
\end{align}
Then  (\ref{215})  may be written as 
\begin{equation}\la{217}
    \begin{aligned}
        &\chi^+_{R}\Big(1-\frac{i}{4}F(e)\, \chi^-_{L}\psi^+_R\Big)=0,\qquad  \qquad \chi^-_{L}\Big(1+\frac{i}{4}F(e)\, \chi^+_{R}\psi^-_L\Big)=0 \ , 
    \end{aligned}
\end{equation}
which is solved by\foot{The expressions in the brackets in \rf{217} are  invertible and thus  setting them to zero does not
give extra   solutions.}
\be \la{218}   \chi^+_{R}=\chi^-_{L}=0   \ .\ee
Combined  with \rf{555}    this gives $\chi_\mu=0$  leading to \rf{L3}. 
Note that  \rf{212}  with $\chi_\m$  determines $\psi_a$  in terms of $\psi^i\del X^i$ 
 but this  has no consequence as $\psi_a$ drops out of the action. 

The next step  is to eliminate from \rf{L3}   the  remaining  two   auxiliary  fields 
represented by $e^a_\mu$ subject to the gauge  conditions \rf{ge}. 
This can be done  by taking variation of \rf{L3}  over $e^a_\mu$ 
 constrained by \rf{ge} (i.e. adding the corresponding projectors),  
 solving  for $e^a_\mu$  in terms of $(X^i, \psi^i)$ and plugging the result back into
  the Lagrangian \rf{L3}.
  Equivalently, we may  add the gauge   conditions \rf{ge}  to \rf{L3}
with Lagrange multipliers   and vary unconstrained $e^a_\mu$.


The resulting   variation  over the zweibein  leads to the condition of the 
 vanishing of the effective  traceless  and symmetric part of  the stress tensor corresponding to \rf{L3}\foot{An alternative   to imposing  the Lorentz and Weyl gauges \rf{ge} on $e^a_\m$ 
in order to eliminate  it from \rf{L3}  is  to redefine $\psi^i$  so that to make the Lagrangian \rf{L3} 
manifestly Lorentz and Weyl   invariant. Indeed, rescaling  $\psi^i$ by  $e^{-1/4} $ we  find   the fermionic term in \rf{L3} 
taking the form 
$ i  e\,   e^\mu_a \, \bar{\psi}^i \gamma^a \partial_\mu \psi^i 
 \to  i    e^{1/2}    e^\mu_a \psi'^i \gamma^a \partial_\mu \psi'^i  $. The resulting  analog \rf{L3}  is now invariant under   the rescaling of $e^a_\mu$ with $\psi'^i$ not transforming. The   variation  of  the action over $e^a_\m$ is then manifestly traceless (with no need to use  the fermionic  equations of motion). 
 Similarly, we may  Lorentz-rotate $\psi^i$  and  $e_{a\m}$ to make the latter symmetric  so 
 that the variation of  the resulting \rf{L3} over  the zweibein will  also be  manifestly symmetric. 
 We then  end up with the same relation as in \rf{219}.} 
\begin{equation}
\label{219}
    \begin{aligned}
    e^\nu_{(a}\big(\eta_{ \mu)  \nu}+ \partial_{\mu)}X^i \partial_\nu X^i \big) +\frac{i}{2}\bar{\psi}^i \gamma_{(a}\partial_{\mu)}\psi^i
    - \frac{1}{2}e_{a\mu}\Big[ \big(\eta_{\l \nu} + \partial_\l X^i \partial_\nu X^i \big)g^{\l \nu}+
    \frac{i}{2}e^\l_b\,  \bar{\psi}^i \gamma^b \partial_\l \psi^i\Big]=0  \ .   \end{aligned}
\end{equation}
Contracting \rf{219} with $e^a_\nu$, symmetrizing  the indices $\mu$ and $\nu$, and taking the determinant
     allows to rewrite the Lagrangian in \rf{L3}   in the "square-root"  form 
\begin{align}
    \mathcal{L} &=-\sqrt{-\det\Big[\eta_{\mu \nu}+\frac{i}{2}e^a_{\mu} e^\rho_{(a}\partial_{\nu)}X^i \partial_\rho X^i
    +\frac{i}{2}e^a_{\nu}e^\rho_{(a}\partial_{\mu)}X^i \partial_\rho X^i 
    +\frac{i}{4}e^a_{\mu}\bar{\psi}^i \gamma_{(a}\partial_{\nu)}\psi^i
    +\frac{i}{4}e^a_{\nu}\bar{\psi}^i \gamma_{(a}\partial_{\mu)}\psi^i
  \Big]}\no \\
  &\qquad \qquad-\frac{i}{4} e^\m_a \, \bar{\psi}^i \gamma^a \partial_\mu \psi^i \ .  \label{Lg}
\end{align}
Here  $e^a_\m$  should  be replaced  by  its expression in terms of $X^i$ and $\psi^i$  found by   solving  \rf{219}. 

Eq. \rf{219}   is a non-linear  algebraic  equation  on  the matrix $e^a_\mu$ subject  to \rf{ge}. 
It  does not seem  possible  to write its solution in a closed form
but  it is sufficient for our goals  here 
to solve  it    perturbatively  by expanding in   powers of 
 $X^i$ and $\psi^i$, i.e. 
\begin{equation}\la{220}
    e^a_\mu =\delta^a_\mu +{\td e}^a_\mu+...,\qquad \qquad g_{\m\n} = \eta_{\mu \nu} + 2\eta_{a(\mu}{\td e}^a_{\nu)} + ... \ , 
\end{equation}
\begin{equation}\la{221}
    \td e_{a\mu} =\frac{1}{2}\partial_{a}X^{i} \partial_{\mu}X^{i}+\frac{i}{4}\bar{\psi}^i\gamma_{(a}\partial_{\mu)}\psi^{i}
    - {1\ov 4} 
     \eta_{a\mu}\eta^{ \sigma \rho} \Big[\partial_\sigma X^i \partial_\rho X^i+\frac{i}{2}\bar{\psi}^i \gamma_a \delta^a_{(\sigma}\partial_{\rho)}\psi^i\Big] \ , 
\end{equation}
where we  imposed the conditions \rf{ge}.  This  procedure  can be extended to higher orders of expansion of $e^a_\mu$
 and the result should be plugged into \rf{Lg}.


The final step is to expand \rf{Lg} in powers of derivatives of the fluctuations $(X^i,\psi^i)$  near the long string vacuum. 
To compute the  tree-level  and one-loop $2\rightarrow2$  scattering amplitude of  $X^i$   one needs  to  keep only the following quadratic   and quartic terms\foot{Note that there are no cubic $\psi \psi X $ vertices. 
To compute the one-loop  tadpole   contribution  to the amplitude  requires  also the knowledge  of  6-point vertices 
but  like in \ci{Seibold:2023zkz,Seibold:2024oyr}  the  resulting  contribution vanishes in dimensional regularization.}
\begin{equation}
\label{L}
\begin{aligned}
    \mathcal{L}
     =&-\frac{1}{2}\partial_{a}X^{i}\partial^{a}X^{i} -{i}\bar{\psi}^{i}\gamma^{a}\partial_{a}\psi^{i} \\ &
      +\frac{1}{4}\partial_{a}X^{i}\partial_{b}X^{i}\partial^{a}X^{j}\partial^{b}X^{j}
  -\frac{1}{8}\partial_{a}X^{i}\partial^{a}X^{i}\partial_{b}X^{j}\partial^{b}X^{j}
\\ & +\frac{i}{2}\partial_{a}X^{i}\partial_{b}X^{i}\bar{\psi}^{j}\gamma^{a}\partial^{b}\psi^{j}
 -\frac{i}{4}\partial_{a}X^{i}\partial^{a}X^{i}\bar{\psi}^{j}\gamma^{b}\partial_{b}\psi^{j}
  \\ &-\frac{1}{4}\bar{\psi}^{i}\gamma_{a}\partial_{b}\psi^{i}\bar{\psi}^{j}\gamma^{(a}\partial^{b)}\psi^{j} +\frac{1}{8}\bar{\psi}^{i}\gamma^{a}\partial_{{a}
 }\psi^{i}\bar{\psi}^{j}\gamma^{b}\partial_{b}\psi^{j} + ....
\end{aligned}
\end{equation}
 Here we  dropped the -1 constant, rescaled  $\psi\rightarrow\sqrt{2}\psi$ 
  and  used $a,b$ for the flat-space indices. 
  

\section{Other similar   actions} 

\subsection{Heterotic  string}

The  covariant  heterotic string  action \ci{Gross:1985fr} 
 can be found  by  coupling $(1,0)$  supersymmetric  2d scalar multiplet $(X^A, \psi^A)$ (with $\psi^A$  being  left MW  spinors)
 to  the corresponding   supergravity   multiplet $(e^a_\m, \chi_\mu$)  (with $\chi_\m$  given by  right MW spinors)
 and   adding  also $N=32$   right MW   spinors $\vp^\rr$ ($\rr=1,2, ..., N$)   coupled  to $e^a_\mu$. 
 The resulting  Lagrangian  can be found  from \rf{1.2}   by adding  the chiral projectors on $\psi^A$ and $\chi_\mu$ 
 and adding the kinetic term for $\vp^\rr$, i.e. 
  \begin{align}
\label{122}
\mathcal{L}= &  e\Big[-\frac{1}{2}\partial_{\mu}X^{A}\partial^{\mu}X_{A}-\frac{i}{2}\bar{\psi}^{A}\gamma^{\mu}\partial_{\mu}\psi_{A}+\frac{i}{2}\bar{\chi}_{\mu}\gamma^{\nu}\gamma^{\mu}\psi^{A}\partial_{\nu}X_{A}
-\frac{i}{2}\bar{\vp}^{\rr}\gamma^{\mu}\partial_{\mu}\vp^{\rr}\Big], \\
  & P_+ \psi^A=\psi^A, \ \qquad P_-\chi_{\mu}=\chi_\mu,\qquad   P_- \vp^\rr=\vp^\rr, \qquad 
  P_\pm\equiv \frac{1}{2}(1\pm \gamma^*) \ . \la{222}
\end{align}
In the $\g^a$-matrix  representation we are using the  chiral fermions  will have  one of  the two  spinor components 
set to zero, $\psi^A_R=0$, $\chi_{L\m}=0$, $\vp^\rr_L=0$. The corresponding action is invariant under local 
$(1,0)$ supersymmetry, i.e.  \rf{23}   with the parameter  subject to 
 $P_- \varepsilon =\varepsilon$, i.e. $\varepsilon_L=0$ (see, e.g.,  \ci{Tanii:1985cp}).

To expand \rf{122}  in the static gauge   we may fix the  supersymmetry and superconformal gauges  as in \rf{25},\rf{28}. The   absence 
of the quadratic gravitino term in \rf{122}  simplifies   its elimination  from the action   and we 
end up with the following analog of \rf{L3} 
\begin{align}\label{L4}
    & \mathcal{L}=-\frac{1}{2} e\Big[  (\eta_{\mu \nu}+  \partial_\mu X^i \partial_\nu X^i ) g^{\mu \nu}
     +      i e^\m_a \bar{\Psi}^{\I}\gamma^a \partial_{\m}\Psi^{\I} 
           \ \Big] \ , \\
         & \bar{\Psi}^{\I}\gamma^a \partial_{\m}\Psi^{\I} \equiv    \bar{\psi}^i P_- \gamma^a \partial_\m \psi^i  +  
           \bar{\vp}^\rr P_+\gamma^a \partial_\m  \vp^\rr  \ . \la{34}
\end{align}
To integrate   out $e^a_\m$ we may   fix  again   the Lorentz and Weyl gauges  as in \rf{ge} 
 ending with a direct analogs of the equation 
\rf{219}  and   the Lagrangian   \rf{Lg}.  Expanding it in powers of derivatives of all fields we get 
the following counterpart of \rf{L}
\begin{equation}
\label{LH}
\begin{aligned}
    \mathcal{L}
     =&-\frac{1}{2}\partial_{a}X^{i}\partial^{a}X^{i} -{i}\bar{\Psi}^{\I} \gamma^{a}\partial_{a}\Psi^{\I}  
        \\ &
      +\frac{1}{4}\partial_{a}X^{i}\partial_{b}X^{i}\partial^{a}X^{j}\partial^{b}X^{j}
  -\frac{1}{8}\partial_{a}X^{i}\partial^{a}X^{i}\partial_{b}X^{j}\partial^{b}X^{j}
\\ & +\frac{i}{2}\partial_{a}X^{i}\partial_{b}X^{i}\bar{\Psi}^{\I}\gamma^{a}\partial^{b}\Psi^{\I}
 -\frac{i}{4}\partial_{a}X^{i}\partial^{a}X^{i}\bar{\Psi}^{\I}\gamma^{b}\partial_{b}\Psi^{\I}
  \\ &-\frac{1}{4}\bar{\Psi}^{\I}\gamma_{a}\partial_{b}\Psi^{\I}\bar{\Psi}^{\j}\gamma^{(a}\partial^{b)}\Psi^{\j} +\frac{1}{8}\bar{\Psi}^{\I}\gamma^{a}\partial_{{a}
 }\Psi^{\I}\bar{\Psi}^{\j}\gamma^{b}\partial_{b}\Psi^{\j} + ...\ . 
\end{aligned}
\end{equation}

\subsection{GS  superstring}

It is instructive also  to compare \rf{L} to a similar expansion of the GS action near the long string vacuum
 \ci{Seibold:2024oyr}. Solving for the independent metric $g_{\m\n}$ the type IIA GS  superstring 
  Lagrangian may be written as the sum of the "volume"  and WZ  terms \ci{Green:1983wt}
\begin{align}
     & L= L_1 + L_2 \ , \qquad \qquad L_{1}=-\sqrt{- g},\\
 & L_{2}=-{i}\epsilon^{\mu\nu}\partial_{\mu}X^{A}\Big[\bar{\theta}^{1}\Gamma_{A}\partial_{\nu}\theta^{1}-\bar{\theta}^{2}\Gamma_{A}\partial_{\nu}\theta^{2}\Big]+\epsilon^{\mu\nu}\bar{\theta}^{1}\Gamma^{A}\partial_{\mu}\theta^{1}\bar{\theta}^{2}\Gamma_{A}\partial_{\nu}\theta^{2},
\\
 & g_{\mu\nu}=\Pi_{\mu}^{A}\Pi_{A\nu} \ , \qquad \qquad 
 \Pi_{\mu}^{A}=\partial_{\mu}X^{A}-i\bar{\theta}^{I}\Gamma^{A}\partial_{\mu}\theta^{I},
\end{align}
where   $\theta^{I}$ ($I=1,2$)   are the   10d MW spinors of opposite chirality. 
Fixing the static gauge $X^a=\xi^a$   and expanding in derivatives  we get
    \begin{align}
L_{1}  =&-\sqrt{-\det\Big[\eta_{\mu\nu}+\partial_{\mu}X^{i}\partial_{\nu}X^{i}-2i\bar{\theta}^{I}\Gamma_{(\underline{\mu}}\partial_{\nu)}\theta^{I}-\bar{\theta}^I \Gamma_{\underline{\mu}}\partial_\nu\theta^I \bar{\theta}^J \Gamma^{\underline{\mu}}\partial^{\underline{\nu}}\theta^J+...\Big]}\no \\
  =&-1 -\frac{1}{2}\partial_{\mu}X^{i}\partial^{\underline{\mu}}X^{i}+i\bar{\theta}^{I}\Gamma^{\mu}\partial_{\underline{\mu}}\theta^{I}\no \\
  & +\frac{1}{4}\partial_{\mu}X^{i}\partial_{\nu}X^{i}\partial^{\underline{\mu}}X^{j}\partial^{\underline{\nu}}X^{j}
  -i\bar{\theta}^{I}\Gamma_{\underline{\mu}}\partial_{\nu}\theta^{I}\partial^{\underline{\mu}}X^{j}\partial^{\underline{\nu}}X^{j}-\frac{1}{2}\bar{\theta}^{I}\Gamma_{\underline{\nu}}\partial_{\mu}\theta^{I}\bar{\theta}^{J}\Gamma^{\underline{\mu}}\partial^{\underline{\nu}}\theta^{J}\no \\
& -\frac{1}{8}\partial_{\mu}X^{i}\partial^{\underline{\mu}}X^{i}\partial_{\nu}X^{j}\partial^{\underline{\nu}}X^{j}+\frac{i}{2}\bar{\theta}^{I}\Gamma^{\underline{\mu}}\partial_{{\mu}}\theta^{I}\partial_{{\nu}}X^{j}\partial^{\underline{\nu}}X^{j}+\frac{1}{2}\bar{\theta}^{I}\Gamma^{\underline{\mu}}\partial_{\mu}\theta^{I}\bar{\theta}^{J}\Gamma^{\underline{\nu}}\partial_{\nu}\theta^{J}+...\ ,\la{2288}\\
\label{GSL2}
L_{2} =&-{i}\Big[\bar{\theta}^{1}\Gamma^{*}\Gamma^{\underline{\nu}}\partial_{\nu}\theta^{1}-\bar{\theta}^{2}\Gamma^{*}\Gamma^{\underline{\nu}}\partial_{\nu}\theta^{2}\Big]+\epsilon^{\mu \nu}\bar{\theta}^1 \Gamma^{\underline{\rho}}\partial_\mu \theta^1 \bar{\theta}^2 \Gamma_{\underline{\rho}}\partial_\nu \theta^2+...\ ,
\end{align}
Defining the 
 projectors $  \mathcal{P}_\pm =\frac{1}{2}(1\pm \Gamma^*)$ we can  fix the $\kappa$-symmetry gauge as  \ci{Seibold:2024oyr}
\begin{equation}
\label{230}
    \mathcal{P}_{-} \theta^1=\mathcal{P}_{+} \theta^2=0,\qquad {i.e.} \qquad  \mathcal{P}_{+} \theta^1=\theta^1,\quad \mathcal{P}_-\theta^2=\theta^2.
\end{equation}
Then the expansion  does not  contain   cubic $\theta\theta X$ vertices 
and we get (omitting the explicit factors of projectors in \rf{230})\foot{Adding $L_2$ to $L_1$   gives  gives the standard kinetic terms  for $\theta^{1}$ and $\theta^2$  with extra 
$2P_\pm$ factors. The expression below  is found 
 after rescaling $\theta^{I}\rightarrow {1 \ov \sqrt{2}}\theta^{I}$ and omitting the  explicit  projector  factors. 
 We also   use the  flat-space indices as in \rf{L}.   We thank M. Beccaria and R. Roiban for 
 correcting the coefficients of the  $\theta^4$ terms. 
  }
    \begin{align}
L   =&-\frac{1}{2}\partial_{a}X^{i}\partial^{{a}}X^{i}+i\bar{\theta}^{I}\Gamma^{a}\partial_{{a}}\theta^{I}\no \\
 & +\frac{1}{4}\partial_{a}X^{i}\partial_{b}X^{i}\partial^{{b}}X^{j}\partial^{{a}}X^{j}
 - \frac{1}{8}
 \partial_{a}X^{i}\partial^{{a}}X^{i}\partial_{b}X^{j}\partial^{{b}}X^{j}\no 
 \\
 &
  -   \frac{i}{2}
  \partial_{{a}}X^{i}\partial_{{b}} X^i \, \bar{\theta}^{I}\Gamma^{{a}}\partial^{b}\theta^{I}
   +
   \frac{i}{4}
   \partial_{{b}}X^{i}\partial^{{b}}X^{i} \, \bar{\theta}^{I}\Gamma^{{a}}\partial_{{a}}\theta^{I}  \no   \\
   &
   -  \frac{1}{4} \bar{\theta}^{I} \Gamma_{{a}}\partial_{b}
   \theta^{I}\bar{\theta}^{J}\Gamma^{{b}}\partial^{{a}}\theta^{J}
   +
  \frac{1}{4} \bar{\theta}^{I}\Gamma^{{a}}\partial_{a}\theta^{I}\bar{\theta}^{J}\Gamma^{{b}}\partial_{b}\theta^{J}+...\ . \la{LL}
\end{align}
Under the projection \rf{230} $\theta^{I}$  have in total 16 real anticommuting  components.
Thus we get the same number of physical degrees of freedom and the same  structure of the quartic 
interaction terms  as in the spinning string case in \rf{L}.\foot{
Note that the   signs of the fermionic  bilinear  terms in 
\rf{L}   in \rf{LL}  are  opposite but they can be  made the same  by
  changing a  representation of the gamma  matrices.}
  
  Note that \rf{LL}   contains only $\Gamma_a$  matrices with 2d   indices  and thus one  can chose  their
  representation  so that to  convert $\theta^I$ into  a collection of  eight 2d spinors (in particular,  
  $\bar \theta ^I \Gamma^a \partial_b \theta ^I $   written explicitly  in components 
   becomes the same as  $\bar \psi^i \gamma^a \partial_b \psi^i$). 
   Redefining  $\psi^i$  in \rf{L}  as 
  \begin{equation}
      \psi^i \rightarrow \psi^i-\frac{i}{16} \big[
      \bar{\psi}^j  \gamma^a \partial_a \psi^j  
      + (\bar{\psi}^j \gamma^*   \gamma^a \partial_a \psi^j ) \gamma^* \big]  \,  \psi^i+\dots,
  \end{equation}
  one can establish direct equivalence between the NSR action \rf{L} and  the GS one in \rf{LL}. 
  This shows  that the  correspondence   between the two actions familiar in the light-cone gauge   holds also 
  in the  static gauge.

\subsection{${T\bar{T}}$ deformation of free scalar multiplet}

Given that  the $(\del X)^4$ terms in the expansion of the Nambu action in the static gauge  can be 
viewed  \cite{Cavaglia:2016oda}  as the $T\bar{T}$ deformation \cite{Zamolodchikov:2004ce}
  of the free massless  scalar   theory 
 one  may wonder if  a   similar   interpretation   may apply also  to the spinning string  Lagrangian in \rf{L}. 
 Starting with the free  Lagrangian for $D-2$ scalar multiplets  on a flat 2d background 
 \begin{equation}\la{111}
\mathcal{L}_{0}=-\frac{1}{2}\partial_{a}X^{i}\partial^{a}X^{i}-i\bar{\psi}^{i}\gamma^{a}\partial_{a}\psi_{i}\ ,
\end{equation}
the   traceless symmetric  part of  the corresponding stress tensor is\foot{As usual, the symmetry 
and tracelessness of the stress  tensor  corresponding to \rf{111} 
 hold upon the use of  the  fermionic equations   of motion.  Here we are considering an 
 off-shell deformation by the symmetric  traceless part  only,  but a deformation by the full 
 stress tensor  should be related  to the  one  below by a   redefinition of  the $\psi^i$ field.}
\begin{equation}
\label{314}
T_{ab}=\partial_{a}X^{i}\partial_{b}X^{i}+i\bar{\psi}^{i}\gamma_{(a}\partial_{b)} \psi^{i}- \frac{1}{2}\eta_{ab} \big(\partial_{c}X^{i}\partial^{c}X^{i}+i\bar{\psi}^{i}\gamma^{c}\partial_{c}\psi^{i}\big) \ .
\end{equation}
 Then the leading order of the  $T\bar{T}$ deformation  ($ {\del \mathcal{L} \ov \del \lambda} =-\ha  \det T_{ab}$)   is 
 proportional to  
\begin{equation}\la{237}
\begin{aligned}
      \mathcal{L}_1=-\ha  \det T_{ab} = -\frac{1}{4}(T^a _a T^b _b -T^a _b T^b _a)=&\ \  \frac{1}{4} \partial_a X^i \partial_b X^i \partial^a X^j \partial^b X^j        -\frac{1}{8} \partial_a X^i \partial^a X^i \partial_b  X^j \partial^b X^j
      \\
& +{i\ov 2} \partial_a X^i \partial_b X^i \bar{\psi}^j \gamma^{a} \partial^{b} \psi^j -\frac{i}{4} \partial_a X^i \partial^a X^i \bar{\psi}^j \gamma^b \partial_b \psi^j\\
&-\frac {1}{4}\bar{\psi }^i \gamma_a \partial _b \psi ^i \bar{\psi}^j \gamma^{(a} \partial ^{b )}\psi ^j + \ \frac {1}{8}\bar{\psi }^i \gamma^a \partial _a \psi ^i \bar{\psi}^j \gamma^b \partial _b \psi ^j \ . 
\end{aligned}
\end{equation}
The sum $\mathcal{L}= \mathcal{L}_{0} + \mathcal{L}_{1}$  is thus   equivalent to  \rf{L}.
The corresponding   action   is invariant  under the modified global supersymmetry relating 
$X^i$ and $\psi^i$ as   discussed in Appendix B.\foot{Let us
 note that the  expansion  of the GS action in  \rf{LL}  can also be reproduced as a similar 
 $T\bar{T}$ deformation as in \rf{237}  if  $T_{ab} $    is chosen as the  canonical stress tensor
 corresponding to the quadratic part of \rf{LL}.}

\section{One-loop scattering amplitude}

Our aim  is to compute the 4-scalar  scattering amplitude  corresponding to the Lagrangian \rf{L}. 
We will follow closely the discussion  of the bosonic   and GS cases in \ci{Dubovsky:2012sh,Seibold:2023zkz,Seibold:2024oyr}.
The  amplitude can be written as
\begin{equation}\la{01}
    \mathcal{M}_{i j, k l}=A\, \delta_{i j} \delta_{k l}+B\, \delta_{i k} \delta_{j l}+C\,\delta_{i l} \delta_{j k}\ , 
\end{equation}
where $A, B$ and $C$ are the  annihilation, transmission and reflection coefficients  that are functions of $s,t,u$
(see Appendix A for conventions). These   can be computed in the inverse string  tension $T^{-1} ={ 2\pi \a'}$ expansion, e.g., 
$A= A^{(0)} + T^{-1} A^{(1)} + ...$. 

\begin{figure}[h]
\begin{center}
     \begin{tikzpicture}[scale=0.5]
\begin{scope}[decoration={
    markings,
    mark=at position 0.5 with {\arrow{latex}}}
    ]
\draw[postaction={decorate}] (-1,1) node[left] {$(p_1,i)$}--(0,0);
\draw[postaction={decorate}] (0,0) -- (1,1) node[right] {$(p_3,k)$};
\draw[postaction={decorate}] (0,0) -- (1,-1) node[right]{$(p_4,l)$};
\draw[postaction={decorate}] (-1,-1) node[left] {$(p_2,j)$}--(0,0);
\fill (0,0)  circle (0.1);
\end{scope}
\end{tikzpicture} 
\end{center}
\begin{center}
\begin{tikzpicture}[scale=0.5]
\begin{scope}[decoration={
    markings,
    mark=at position 0.5 with {\arrow{latex}}}
    ]
\fill (1,0)  circle (0.1);
\fill (-1,0)  circle (0.1);
\draw[postaction={decorate}] (-2,2) node[left]{$(p_1,i)$}--(-1,0) ;
\draw[postaction={decorate}] (-2,-2) node[left]{$(p_2,j)$}--(-1,0) ;
\draw[postaction={decorate}]  (1,0)--(2,2)node[right]{$(p_3,k)$} ;
\draw[postaction={decorate}] (1,0)--(2,-2) node[right]{$(p_4,l)$} ;
\draw[postaction={decorate}] (-1,0) .. controls (0,1) .. (1,0) ;
\draw[postaction={decorate}] (-1,0) .. controls (0,-1) .. (1,0) ;
\draw[] (0,-3) node[] {$s$-channel};
\node[anchor=west] at (-1.2,1.5) {$(p,a)$};
\node[anchor=west] at (-1.2,-1.5) {$(q,b)$};
\end{scope}
\end{tikzpicture}
\begin{tikzpicture}[scale=0.5]
\begin{scope}[decoration={
    markings,
    mark=at position 0.5 with {\arrow{latex}}}
    ]
\fill (0,1)  circle (0.1);
\fill (0,-1)  circle (0.1);
\draw[postaction={decorate}] (-2,2) node[left]{$(p_1,i)$}--(0,1) ;
\draw[postaction={decorate}] (-2,-2) node[left]{$(p_2,j)$}--(0,-1) ;
\draw[postaction={decorate}]  (0,1)--(2,2)node[right]{$(p_3,k)$} ;
\draw[postaction={decorate}] (0,-1)--(2,-2) node[right]{$(p_4,l)$} ;
\draw[postaction={decorate}] (0,1) .. controls (1,0) .. (0,-1) ;
\draw[postaction={decorate}] (0,1) .. controls (-1,0) .. (0,-1) ;
\draw[] (0,-3) node[] {$t$-channel};
\node[anchor=west] at (-3,0) {$(p,a)$};
\node[anchor=east] at (3,0) {$(q,b)$};
\end{scope}
\end{tikzpicture}
\begin{tikzpicture}[scale=0.5]
\begin{scope}[decoration={
    markings,
    mark=at position 0.6 with {\arrow{latex}}}
    ]
\fill (0,1)  circle (0.1);
\fill (0,-1)  circle (0.1);
\draw[postaction={decorate}] (-2,2) node[left]{$(p_1,i)$}--(0,1) ;
\draw[postaction={decorate}] (-2,-2) node[left]{$(p_2,j)$}--(0,-1) ;
\draw[postaction={decorate}]  (0,1)..controls(3,1)..(3.5,-2)node[right]{$(p_4,k)$} ;
\draw[postaction={decorate}] (0,-1)..controls(3,-1)..(3.5,2) node[right]{$(p_3,l)$} ;
\draw[postaction={decorate}] (0,1) .. controls (1,0) .. (0,-1) ;
\draw[postaction={decorate}] (0,1) .. controls (-1,0) .. (0,-1) ;
\draw[] (1,-3) node[] {$u$-channel};
\node[anchor=west] at (-3,0) {$(p,a)$};
\node[anchor=east] at (2.9,0) {$(q,b)$};
\end{scope}
\end{tikzpicture}
\end{center}
\caption{\small{Diagrams that contribute to tree-level and one-loop $2 \rightarrow\, $2 bosonic scattering amplitude. }}
    \label{fig1}
\end{figure}
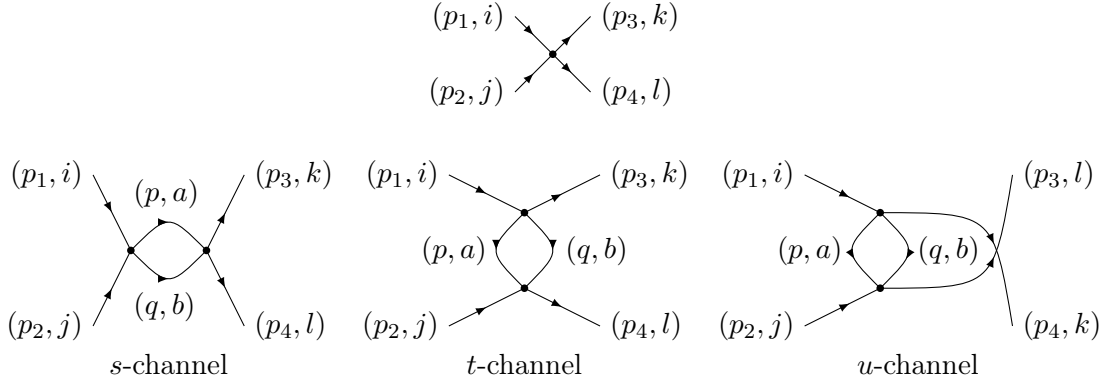

 The  tree-level and one-loop diagrams are  shown  in 
 Figure \ref{fig1}. 
 
The tree-level  amplitude  is the same  as in the   Nambu action   case, i.e. is found from the  $(\del X)^4$
terms in \rf{L}: 
\begin{equation}\la{02}
     A^{(0)}=-\frac{1}{2} t u, \qquad B^{(0)}=-\frac{1}{2} s u, \qquad C^{(0)}=-\frac{1}{2} s t .
\end{equation}
The  massless 2d kinematics  implies $stu=0$   so that choosing $t=0, \ u=-s$ we get 
$A^{(0)}=C^{(0)}=0$, $B^{(0)}=-\frac{1}{2}s^2$. 

The bosonic loop contribution to the amplitude  found in \cite{Dubovsky:2012sh,Seibold:2023zkz} is given by 
(assuming  dimensional regularization $d=2-2\epsilon$  and setting  $\hat D\equiv D-2$)
\begin{align}\la{03}
A_{b}^{(1)}= &-\frac{\hat D -6}{96\pi} stu \Big(\frac{1}{\epsilon}-\gamma+\log4\pi\Big) \no \\
 &-\frac{1}{192\pi}\Big[   (\hat D -24) s^{3}+\big( \frac{16}{3}\hat D  + 12 - 2 (\hat D-6) \log {-s\ov \mu^2} \big)  stu
 + 12 \big( t \log {s\ov t}  + u \log {s\ov u} \big) \Big] \ , 
\end{align}
with $B_{b}^{(1)}$ and $C_{b}^{(1)}$   obtained  by interchanging $s,t,u$. 
Specifying  to 2d kinematics  with $t=0$   gives 
\begin{equation}\la{340}
A_b^{(1)}=- C_b^{(1)}= -\frac{{\hat D}-24}{192 \pi} s^3\ ,\ \ \ \  \ \ \qquad B_b^{(1)}=\frac{i}{16} s^3,\ \ 
\end{equation}
so that for  $D=26$  one has $A^{(1)}_b=C^{(1)}_b=0$ and $B^{(1)}_b=\frac{i}{16}s^3$, in agreement with the integrability of underlying  critical string theory. 

The $\del X \del X \psi \del \psi$   vertex in \rf{L}  implies  the presence of the  {fermionic loop} contribution
  to the amplitude.
The  corresponding vertex multiplying $\delta_{ij}\delta_{kl}$   in momentum space may be written as 
\begin{equation}
\begin{aligned}
    V(p_1,p_2,p_3,p_4)^\alpha_\beta=&\frac{1}{2}\Big[ \;(p_1\cdot p_2)\;(p_3\cdot \gamma)^\alpha_\beta-(p_1\cdot p_3)\;(p_2\cdot\gamma)^\alpha_\beta-  (p_2\cdot p_3)\;(p_1\cdot\gamma)^\alpha_\beta\Big]\ , \\
\end{aligned}
\end{equation}
where $\alpha,\beta=1,2$ are 2d spinor indices. 
  For example, the  contribution to the $A$-part of the amplitude  in the $s$-channel is 
  ($q=p_1+p_2-p$)\foot{In computing the amplitude with a  loop of  Majorana fermions  we follow the standard  prescription
(see, e.g., \ci{Denner:1992me}), i.e. do computation for Dirac  fermions and multiply the result by 
 $\ha$.  We also drop the contribution of the second $\del X \del X \psi \del \psi$  term in \rf{L} that  vanishes 
  on the fermionic equation of motion which  does not contribute.  It can be eliminated by a redefinition of $\psi$, or, 
  equivalently, the corresponding  contribution  to  the fermionic loop 
   reduces to a tadpole diagram that  vanishes  in dimensional regularization.}
 \begin{align}
\label{intN}
& A^{(1)}_{f,s}=  \frac{1}{2i}  \int \frac{d^d p}{(2\pi)^d}\frac{N_{A,s}}{(p^2-i\varepsilon)(q^2-i\varepsilon)}=\frac{1}{2i}\int^1_0 dx \int \frac{d^d l}{(4\pi)^d}\frac{N_{A,s}}{(l^2+\Delta)^2},\\
    & N_{A,s}=\hat D\ \Tr \Big[ V(p_1,p_2,p,q)\  (p\cdot \gamma)\  V(-p_3,-p_4,q,p)\ (q\cdot \gamma)\Big]\ , \\
     &p=l+x(p_1+p_2),\qquad q=(1-x)(p_1+p_2)-l,\qquad \Delta=-x(1-x)s -i\varepsilon\ .  \la{37}
\end{align}
To evaluate the integral,  $N_{A,s}$ can be decomposed in terms of powers of $l^a$  as 
\begin{align}
&\label{Ns}
N_{A,s}=N_0+N_2 l^2+N_{a b} l^a l^b+N_4(l^2)^2+N_{a b c e } l^a l^b l^c l^e+M_{a b} l^2 l^a l^b,\\
   & N_0=-\frac{1}{16} n_f\, \hat{D}\,s^2(t+u)^2(-1+x)^2x^2,\quad N_2=-\frac{1}{16} n_f\, \hat{D}\,s(t+u)^2(-1+x)x, \quad N_4=0\ , \no \\
   &N_{ab}=\frac{1}{16} n_f\, \hat{D}\,s(1-x)(mp_{1a }p_{3b}+mp_{2a}p_{4b}+np_{2a}p_{3b}+np_{1a}p_{4b}),\no \\
   &M_{ab}=\frac{1}{8}n_f\, \hat{D}\, (up_{2a}p_{3b}+up_{1a}p_{4b}+tp_{2_a}p_{4b}+tp_{1a}p_{3b}),\qquad N_{abce}=-2 n_f\, \hat{D}\,p_{1a}p_{2b}p_{3c}p_{4e} ,\no \\
   &\qquad m=s+2(u-2s)x,\quad n=s+2(t-2s)x\ . \no
\end{align}
Here $\hat D=D-2$  is the number of vector  components  of $\psi^i$ running in the loop   and  $n_f=2$  is  the number of 2d spinor components.
In 
dimensional regularization ($d=2-2\epsilon$)  the integral (\ref{intN}) can be written as
\begin{align}
    & A^{(1)}_{f,s}=\frac{1}{4\pi}\int^1_0 dx \Big[ K_1 \big(\frac{1}{\epsilon}-\gamma +\ln 4\pi\big) -K_1 \ln \Delta +K_0\Big]\ , 
\\
    &\quad K_1 = N_2 +\frac{1}{2}N_{a b}\eta^{a b}-2N_4 \Delta -\Delta M_{a b}\eta^{a b}-\frac{1}{4}\Delta N_{a b c e}(\eta^{a b}\eta^{c e}+\eta^{a c}\eta^{b e}+\eta^{a e}\eta^{b c}),\\
    &\quad K_0=\frac{1}{\Delta}N_0 -N_2 +\Delta N_4 -\frac{1}{2}\Delta M_{a b}\eta^{a b}-\frac{1}{4}\Delta N_{a b c e}(\eta^{a b}\eta^{c e}+\eta^{a c}\eta^{b e}+\eta^{a e}\eta^{b c}) \ , 
\end{align}
where $\Delta$ was defined in \rf{37}. 
As a  result, the fermionic loop contribution to the  $s$-channel $A$-amplitude    can be written   as 
\begin{equation}\la{314}
    A^{(1)}_{f,s}=-\frac{1}{2}\frac{ n_f\, \hat{D}\,stu}{192\pi}\Big(\frac{1}{\epsilon}-\gamma+\log4\pi\Big)-\frac{1}{4}\frac{n_{f}\hat{D}}{192\pi}\Big(s^{3}+\frac{4}{3}stu-2stu\log\frac{-s}{\m^{2}}\Big).
\end{equation}
Similarly, we get  $A^{(1)}_{f,t}=A^{(1)}_{f,u}=0$  so that the  total fermionic  loop  contributions are 
\begin{align}
 & A_{f}^{(1)}=-\frac{1}{2}\frac{ n_f\, \hat{D}\,}{192\pi}stu \Big(\frac{1}{\epsilon}-\gamma+\log4\pi\Big)-\frac{1}{4}\frac{n_{f}\hat{D}}{192\pi}\Big[s^{3}+\frac{4}{3}stu-2stu\log\frac{-s}{\mu^{2}}\Big],\la{315}\\
 & B_{f}^{(1)}=-\frac{1}{2}\frac{ n_f\, \hat{D}\,}{192\pi}stu\Big(\frac{1}{\epsilon}-\gamma+\log4\pi\Big)-\frac{1}{4}\frac{n_{f}\hat{D}}{192\pi}\Big[t^{3}+\frac{4}{3}stu-2stu\log\frac{-t}{\mu^{2}}\Big],\la{316}\\
 & C_{f}^{(1)}=-\frac{1}{2}\frac{ n_f\, \hat{D}\,}{192\pi}stu\Big(\frac{1}{\epsilon}-\gamma+\log4\pi\Big)-\frac{1}{4}\frac{n_{f}\hat{D}}{192\pi}\Big[u^{3}+\frac{4}{3}stu-2stu\log\frac{-u}{\mu^{2}}\Big].\la{317} 
\end{align}
For   $t=0$   this  gives 
\be \la{101}
A_{f}^{(1)}=- C_{f}^{(1)}=  -\frac{1}{4}\frac{ n_f\,\hat{D}}{192\pi}\ s^{3}\ , \qquad \qquad B_{f}^{(1)}=0 \ . \ee 
Summing together the 
bosonic \rf{03} and the  fermionic \rf{315} loop   contributions we  find  for 
 the total one-loop  $A$ amplitude 
\begin{align}
    A^{(1)}  =&- \frac{(1 + {1\ov 4} n_f)\hat{D}-6}{ 96\pi}stu\Big(\frac{1}{\epsilon}-\gamma+\log4\pi\Big)\no \\
 & -\frac{1}{192\pi}\Big\{\Big[(1+\tfrac{1}{4}n_f)\, \hat{D}-24\Big]s^{3} + \Big(\big(\tfrac{16}{3} + \tfrac{1}{3} n_f\big) \hat{D}+12-2\big[(1+\tfrac{1}{4}n_f)\hat{D}-6\big]\log\frac{-s}{\mu^{2}}\Big)stu\no \\
&\qquad \qquad \ \ +12\Big(t\log\frac{s}{t}+u\log\frac{s}{u}\Big)tu
 \Big\}\ ,  \la{318}
\end{align}
 with $B^{(1)}$ and $C^{(1)}$  given by interchanging $s,t,u$  as in \rf{315}--\rf{317}.
 
 Specifying to the  2d kinematics  with $t=0$ we get   the  sum of  \rf{340}  and \rf{101}, i.e.
 \be \la{319}
A^{(1)}=- C^{(1)}= -\frac{1}{192 \pi}  \big[ (1 + \tfrac{1} {4} n_f)\hat{D}-24 \big]    s^3,
 \ \ \qquad B^{(1)}=\frac{i}{16} s^3 \ . 
\ee
 In the    critical spinning string case where $\hat D=8$ and $n_f=2$ 
 we thus  finish with 
  \begin{align}\la{320} 
A^{(1)}  =-  C^{(1)}= \frac{1}{16\pi}s^{3},\qquad\qquad 
B^{(1)}  =\frac{i}{16}s^{3} \ . 
\end{align}
The  above expressions for the amplitudes \rf{318}  and \rf{319}   with $n_f=2$  are the  same as found for the GS superstring case  \cite{Seibold:2024oyr}.

The imaginary part of the amplitude given   by $B^{(1)}$  does not receive a contribution from the fermionic loop 
(cf. \rf{101}), i.e. it  is the same  as in the bosonic   string case \rf{340}   and thus   is consistent with "pure-phase" structure of the S-matrix   consistent with  integrability \ci{Dubovsky:2012sh}. 
As discussed in \cite{Seibold:2024oyr}, one can  make $A^{(1)}  =-  C^{(1)}= 0$ by adding a local counterterm.

The expected  agreement between the spinning string  and GS superstring amplitudes  
 follows from the close similarity  of the corresponding 4-vertices in the  Lagrangians  in  \rf{L}  and  \rf{LL}. 
In the  GS  case $\theta^I$  subject to \rf{230}   have 16 independent components. 
 In the spinning string case  $\psi^i$  has $n_f\, \hat D =2 (D-2) $ components  which  for $D=10$ gives  the 
 same  factor of 16 in the fermionic  loop.

  In the case of the heterotic string   we find from \rf{LH}
  that the coefficient $q_s= n_f \hat D$  in the fermion loop   contributions \rf{315}--\rf{101} 
  is  replaced by $ q_h=\ha  n_f (  \hat D  + N)$    where $\ha$ is due to the  chiral projection and $N$ is  the number 
  of right internal fermions $\vp^\rr$.  For the critical string  case of  $\hat D=8$ and $n_f=2$ we thus get
  $q_h= 40$  instead of  $q_s= 16$  in the spinning  string and the superstring cases. 
  The coefficient $\hat D   + {1\ov 4} q_s -24= -12 $ in \rf{319} 
  then becomes  $\hat D   + {1\ov 4} q_h -24= - 6 $. 
\section*{Acknowledgements}
 We  thank  F. Seibold for  useful discussions and   sharing a Mathematica code  for computing 
 one-loop amplitudes in the GS  string case. 
AAT is  grateful to R. Metsaev for  important   discussions   and  suggestions.
   ZW would like   to thank H. Jiang, K. Mkrtchyan and  D. Zhong for many discussions. 
   We also thank  R. Metsaev  and F. Seibold for comments on the draft. 
This   work  was  supported   by the STFC grant ST/T000791/1.

\appendix
\section{Conventions   and useful relations}
\label{AA}
For the $2\rightarrow2$ scattering, the two incoming massless scalars are 
 labeled by $i,j$ with momenta $p_1, p_2$ and two outgoing ones   with  $k,l$ with momenta $p_3, p_4$, 
   so that  in 2d case 
\begin{align}
  &  s=-(p_1+p_2)^2= -2p_1\cdot p_2 ,\qquad t=-(p_1-p_3)^2= 2p_1\cdot p_3 ,\qquad u=-(p_1-p_4)^2= 2p_1\cdot p_4\ ,\no  \\
   & \qquad  p_1 +p_2 =p_3 +p_4  \ , \qquad p^2_r=0 \ , \qquad  \ \ \  s+t+u=0 \ , \ \ \ \ \ \ \ \ \    stu=0 \  .
\end{align}
The   $\gamma^a$  matrices   are  chosen as 
\begin{equation}
    \gamma^0=\Big(\begin{array}{cc}
        0 &-1  \\
         1& 0
    \end{array}\Big),
    \qquad \qquad  \gamma^1=\Big(\begin{array}{cc}
        0 &1  \\
         1& 0
    \end{array}\Big),
    \qquad  \qquad \gamma^*=\gamma^0\gamma^1=\Big(\begin{array}{cc}
        -1 &0  \\
         0& 1
    \end{array}\Big).
\end{equation}
The 2d fermions  $\psi $  have  spinor components  $(\psi_{R},\psi_{L})$ ,   with 
$\bar{\psi }=(\psi_{L},-\psi_{R})$. 
We  also define 
\begin{align}
 &   \xi^\pm=\xi^0\pm \xi^1,\qquad \partial_\pm = \frac {1}{2}(\partial_0 \pm \partial_1), \qquad \psi^\pm = \psi^0\pm\psi^1\ , 
\\
&\psi^{a}\psi'_{a}  =-\frac{1}{2}(\psi^+\psi'^-+\psi^-\psi'^+),\qquad\qquad 
\partial^{a}X^{i}\partial_{a}X^{i} =-4\partial_{+}X^{i}\partial_{-}X^{i}.
\end{align}
One finds the following  relations 
\begin{align}\label{g0d}
&\bar{\psi}^{a}\gamma^{0}\partial_{b}\psi_{a}  =-\psi_{L}^{a}\partial_{b}\psi_{a,L}-\psi_{R}^{a}\partial_{b}\psi_{a,R} 
=\frac{1}{2}\psi_{L}^{+}\partial_{b}\psi^-_{L}+\frac{1}{2}\psi_{L}^{-}\partial_{b}\psi^+_{L}+\frac{1}{2}\psi_{R}^{+}\partial_{b}\psi^-_{R}+\frac{1}{2}\psi_{R}^{-}\partial_{b}\psi^+_{R}\\
&\bar{\psi}^{a}\gamma^{1}\partial_{b}\psi_{a} 
 =\psi_{L}^{a}\partial_{b}\psi_{a,L}-\psi_{R}^{a}\partial_{b}\psi_{a,R}
=-\frac{1}{2}\psi_{L}^{+}\partial_{b}\psi^-_{L}-\frac{1}{2}\psi_{L}^{-}\partial_{b}\psi^+_{L}+\frac{1}{2}\psi_{R}^{+}\partial_{b}\psi^-_{R}+\frac{1}{2}\psi_{R}^{-}\partial_{b}\psi^+_{R} \label{g1d} \ , \\
&\bar{\psi}^{a}\gamma^{b}\partial_{b}\psi_{a}  =-2(\psi_{L}^{a}\partial_{-}\psi_{a,L}+\psi_{R}^{a}\partial_{+}\psi_{a,R}) =\psi_{L}^{+}\partial_{-}\psi^-_{L}+\psi_{L}^{-}\partial_{-}\psi^+_{L}+\psi_{R}^{+}\partial_{+}\psi^-_{R}+\psi_{R}^{-}\partial_{+}\psi^+_{R}\ . 
\end{align}

\iffa 
In computing the amplitude with a  loop of  Majorana fermions  we follow the standard  prescription
(see, e.g., \ci{Denner:1992me}), i.e. do computation for Dirac  fermions and multiply the result by 
 $\ha$.  
 The fermionic contribution to the one-loop $A$-amplitude in \rf{01}  from the $s$-channel  may be written as 
\begin{align}
 &  A^{(1)}_{f,s}=\frac{D-2}{8i}\int_{0}^{1}dx\int\frac{d^{2}p}{(2\pi)^{2}}\frac{1}{(l^{2}+\Delta)^{2}}(p_{4}.p)(p_{2}.q)\, \Tr\Big[(p_{1}.\gamma)(q.\gamma)(p_{3}.\gamma)(p.\gamma)\Big]+...\ , \la{a8}
\end{align}
where $q=p_1+p_2-p,\;l=p-x(p_{1}+p_{2}),\;\Delta\equiv -x(1-x)s-i\varepsilon.$  In \rf{a8}  dots   stand for the contributions 
of  three more diagrams by different Wick contractions  and the trace is over 2d spinor indices. 
\fi
\iffa 
Some standard  one-loop integrals  are
\small {
\begin{align}
& \no I_0 
= \int \frac{d^dp}{(2 \pi)^d} \frac{1}{(p^2 + \Delta)^n} = \frac{i\Gamma(n-\frac{d}{2})}{(4 \pi)^{d/2} \Gamma(n)}  \Big(\frac{1}{\Delta} \Big)^{n-\frac{d}{2}}~,   \quad 
I_2=\int \frac{d^dp}{(2 \pi)^d} \frac{p^2}{(p^2 + \Delta)^n} 
= \frac{i\Gamma(n-1-\frac{d}{2})}{(4 \pi)^{d/2} \Gamma(n)} \Big(\frac{1}{\Delta} \Big)^{n-1-\frac{d}{2}} \frac{d}{2}~,  \\
&\no I_2^{ab}=\int \frac{d^dp}{(2 \pi)^d} \frac{p^\mu p^\nu}{(p^2 + \Delta)^n} =  \frac{i\Gamma(n-1-\frac{d}{2})}{(4 \pi)^{d/2} \Gamma(n)}  \Big(\frac{1}{\Delta} \Big)^{n-1-\frac{d}{2}} \frac{1}{2} \eta^{ab}~, \\ &
\no I_4=\int \frac{d^dp}{(2 \pi)^d} \frac{(p^2)^2}{(p^2 + \Delta)^n} =  \frac{i\Gamma(n-2-\frac{d}{2}) }{(4 \pi)^{d/2} \Gamma(n)} \Big(\frac{1}{\Delta} \Big)^{n-2-\frac{d}{2}}\frac{d(d+2)}{4} ~, \\
&\label{Imnrs}I_4^{abce}=\int \frac{d^dp}{(2 \pi)^d} \frac{p^a p^b p^c p^e}{(p^2 + \Delta)^n} = 
 \frac{i\Gamma(n-2-\frac{d}{2})}{(4 \pi)^{d/2} \Gamma(n)}  \Big(\frac{1}{\Delta} \Big)^{n-2-\frac{d}{2}} \frac{1}{4} \Big(\eta^{ab} \eta^{ce} + \eta^{ac} \eta^{be} + \eta^{a e} \eta^{bc} \Big)~.
\end{align}
}
\fi

\section{Global supersymmetry of   expanded action} 
\label{AB}

The free part of   the Lagrangian \rf{L} or $\mathcal{L}_0$ in \rf{111}  is invariant  under  the  standard global supersymmetry 
(cf. \rf{23}), i.e. 
$  \delta X^i \sim  i \bar{\varepsilon} \psi^i\,, \ \delta \psi^i \sim  \partial X^i\ve$.
Including the   quartic interaction terms in \rf{L} or $\mathcal{L}_1$ in \rf{237} 
one can  check that  it is possible to deform  these transformations by
 extra $\del X \del X \psi\,  \ve + \psi \del \psi \psi\,  \ve$  terms 
so that the total action is again invariant. 

Starting with  $\mathcal{L}_1$   \rf{237}  with generic coefficients $\vv_1,...,\vv_4$ in front of the last  four fermionic terms
 and  writing it in terms of   light-cone derivatives  and  spinor components we get 
\begin{align}
     \mathcal{L}_0 =&2 \partial_{+} X^i \partial_{-} X^i+2 i \psi_L^i \partial_{-} \psi_L^i+2 i \psi_R^i \partial_{+} \psi_R^i\ ,\no  \\
     \mathcal{L}_1 = &2 \partial_{+} X^i \partial_{+} X^i \partial_{-} X^j \partial_{-} X^j+4  \vv_1\partial_{-} X^i \partial_{-} X^i\, \psi_L^j \partial_{+} \psi_L^j\no \\
 &+4  \vv_1\partial_{+} X^i \partial_{+} X^i\, \psi_R^j \partial_{-} \psi_R^j+8  \vv_3\psi_L^i \partial_{+} \psi_L^i\, \psi_R^j \partial_{-} \psi_R^j\la{b1}  \\
& +\left(4  \vv_1+8  \vv_2\right)\, \partial_{+} X^i \partial_{-} X^i\, \big(\psi_L^j \partial_{-} \psi_L^j+\psi_R^j \partial_{+} \psi_R^j\big)+\left(2  \vv_3+4  \vv_4\right)\big(\psi_L^i \partial_{-} \psi_L^i+\psi_R^i \partial_{+} \psi_R^i\big)^2\ . \no 
\end{align}
Considering the  deformation $ \delta=\delta_{(0)}+\delta_{(1)}+...$
of the leading-order supersymmetry  transformations 
\begin{equation}
\label{susy0}
    \delta_{(0)} X^i=i \varepsilon_L \psi_R^i-i \varepsilon_R \psi_L^i, \qquad \delta_{(0)} \psi_R^i=-\partial_{-} X^i \varepsilon_L, \qquad \delta_{(0)} \psi_L^i=\partial_{+} X^i \varepsilon_R \ , \qquad \ \  \delta_{(0)}\mathcal{L}_0=0\ , 
\end{equation}
and demanding the invariance $(\delta_{(0)}+\delta_{(1)}+...) (  \mathcal{L}_0 +  \mathcal{L}_1 + ...) =0$, i.e.  
$
    \delta_{(0)}\mathcal{L}_1+\delta_{(1)}\mathcal{L}_0=0 \ , 
$
one can  fix both  the coefficients $\vv_r$ in \rf{b1} and in  the deformed supersymmetry  transformations  $ \delta_{(1)} X^i,  \delta_{(1)} \psi^i$. 
One  solution is\foot{We assume  trivial boundary conditions on an infinite line, i.e. drop all boundary terms.} 
\begin{align}
 &\qquad \qquad   \vv_1=\frac{i}{2}  ,\qquad  \vv_2=-\frac{i}{4},\qquad \vv_3=-\frac{1}{4} ,\qquad  \vv_4=\frac{1}{8}, \la{b4}
\\
 \delta_{(1)} X^i & =i \partial_{+} X^i \partial_{-} X^j \psi_R^j \varepsilon_L-i \partial_{-} X^i \partial_{+} X^j \psi_L^j \varepsilon_R-\frac{1}{2} \psi_R^j \partial_{+} \psi_R^j \psi_L^i \varepsilon_R+\frac{1}{2} \psi_L^j \partial_{-} \psi_L^j \psi_R^i \varepsilon_L \ , \no \\
\delta_{(1)} \psi_L^i & =(\partial_{+} X^j)^2  \partial_{-} X^i \varepsilon_R+i \partial_{-} X^j \partial_{+} \psi_L^i \psi_R^j \varepsilon_L-i \partial_{-} X^i \partial_{+} \psi_L^j \psi_L^j \varepsilon_R-\frac{i}{2} \partial_{-} X^j \partial_{+} \psi_R^j \psi_L^i \varepsilon_L\ , \no \\
\delta_{(1)} \psi_R^i & =-(\partial_{-} X^j )^2\partial_{+} X^i \varepsilon_L-i \partial_{+} X^j \partial_{-} \psi_R^i \psi_L^j \varepsilon_R+i \partial_{+} X^i \partial_{-} \psi_R^j \psi_R^j \varepsilon_L+\frac{i}{2} \partial_{+} X^j \partial_{-} \psi_L^j \psi_R^i \varepsilon_R  \ . \la{b5}
\end{align}
The  corresponding deformed   supersymmetry algebra closes on the equations of motion.

The values of  the  coefficients  $\vv_r$  in \rf{b4} correspond precisely to the ones in \rf{L} or \rf{237}. 
In general,  the  values of  $\vv_2$  and $\vv_4$   that multiply the terms in $  \mathcal{L}_1$ that are 
proportional to the fermionic equations of motion (i.e. the  fourth and sixth terms in \rf{237}) 
 can be changed  by a local redefinition of $\psi^i$.\foot{Explicitly,  given $\bar \psi \gamma^a \del_a \psi  + f(\xi) \bar \psi \gamma^a \del_a \psi$    (where here  $f\sim \del X \del X + \psi \del \psi $)  we can eliminate the $f$-term  by 
 redefining  $\psi \to \psi'=(1 - \ha f + ...) \psi $.}

\ed

\small 
\bibliographystyle{JHEP-v2.9}
\small
\bibliography{biblio2.bib}
\end{document}